\documentclass[aps,showpacs,twocolumn,superscriptaddres,nofootinbib]{revtex4-1}

\usepackage{graphicx}

\usepackage{epstopdf}
\usepackage{color, amsmath, bm}
\usepackage[dvipsnames]{xcolor}
\usepackage[hidelinks]{hyperref}
\usepackage{natbib}
\usepackage{circuitikz}% \bibliographystyle{abbrvnat}

\usepackage{amsfonts}   
\usepackage{physics} 

\usepackage{subcaption}
\usepackage{caption}
\usepackage{float}

\begin{document}

\title{Dynamically Enabled Robustness of Geometric Phases and Entanglement in the Nonlinear Jaynes–Cummings Model}
\author{Ali Martin Zynda}
\author{Paula I. Villar}
\author{Fernando C. Lombardo}
\affiliation {Departamento de F\'\i sica {\it Juan Jos\'e
Giambiagi}, FCEyN UBA Facultad de Ciencias Exactas y Naturales,
Ciudad Universitaria, Pabell\' on I, 1428 Buenos Aires, Argentina.}
\affiliation{\it Instituto de F\'\i sica de Buenos Aires (IFIBA), CONICET-}

\date{\today}

%====================================================================
\begin{abstract}   
Robustness in dissipative light–matter systems has recently been associated with resonance conditions or geodesic evolution. We show that, in the nonlinear Jaynes–Cummings model, these conditions are necessary but not sufficient. Using a Kerr-type extension together with a Lindblad description of cavity losses and atomic decoherence, we identify a dynamically enabled mechanism in which the stability of geometric phases and entanglement is governed by the alignment between coherent and dissipative trajectories in Hilbert space. Our results reveal that environmental action does not merely suppress quantum features, but reshapes the geometry of state-space evolution: protection emerges only when dissipation preserves the structure of the underlying unitary dynamics. This establishes a general geometric criterion for decoherence resilience in nonlinear light–matter systems and provides guiding principles for engineering protected evolution in open quantum platforms.
\end{abstract}
\maketitle

%====================================================================	

\section{Introduction}

Hybrid light–matter systems are among the most prominent platforms in modern quantum technologies, as they naturally combine long-lived atomic degrees of freedom with highly controllable bosonic modes. In particular, cavity and circuit quantum electrodynamics (QED) enable coherent interactions between few-level systems (qubits) and quantized electromagnetic fields, offering a flexible setting for quantum information processing, quantum simulation, and engineered open quantum dynamics \cite{Blais2021,haroche2006exploring,Kimble_2008,berman1994cavity}. Within these architectures, system parameters can be tuned in situ, which makes it possible to explore in detail the geometric and dynamical aspects of hybrid quantum evolution.

The Jaynes–Cummings (JCM) \cite{jc_original} Hamiltonian provides the basic theoretical framework for describing such systems, even though it was experimemtally confirmed in 1987, it is still relevant today \cite{knight2024,Shore1993,Knight1991}. For example, hybrid entanglement structures are directly relevant for distributed quantum protocols, quantum networking, and cavity-mediated quantum gates \cite{Azuma_2011,Azuma2016DecoherenceOT}.

Since the pioneering works of Berry and Simon \cite{Berry1984,Mukunda1993}, geometric phases (GPs) have been understood as global quantities that depend exclusively on the path traced by a quantum state in projective Hilbert space. Because of this geometric character, they can display a degree of robustness against certain dynamical fluctuations, which has motivated their application to geometric and holonomic quantum computation \cite{Zhang_2023,Zanardi1999,Sjoqvist_2012,Rus2025}. In holonomic approaches, quantum gates are implemented through carefully designed geometric evolutions, and their stability relies on the geometric properties of the trajectory in Hilbert space rather than on precise dynamical timing. For this reason, a detailed understanding of geometric phases in light–matter models such as the JCM is of particular interest.

Cavity and circuit QED platforms are especially well suited for geometric and holonomic control, as tunable detunings and coupling strengths allow for the controlled engineering of effective Hamiltonians \cite{Blais2021}. Another application to trapped ions \cite{Ganesh2025} relates de GP of the ions to their entanglement, and proposes the GP as a sensor for quantum correlations of the system \cite{mbfarias,Lombardo_2021,Martin_Martinez_2013,Mirkin_2021}.
Although the dynamical properties of the JCM have been extensively studied, the impact of resonance conditions on the geometric phase of dissipative hybrid systems remains comparatively less explored.

This work builds upon the results of Viotti et al. \cite{Viotti_2021}, where geometric phases and resonance conditions were analyzed in the Jaynes–Cummings model. By incorporating Kerr nonlinearities and addressing the interplay between initial-state preparation, dissipation, and entanglement dynamics, we go beyond previous analyses by identifying a qualitatively new mechanism governing robustness: while resonance is a necessary condition, it is not sufficient to ensure stability of the geometric phase under environmental action. Instead, robustness is dynamically determined by the extent to which dissipation preserves the geometric structure of the underlying unitary evolution.

While previous studies analyzed geometric phases and entanglement as isolated phenomena, this work identifies a unified dynamical mechanism governing the protection of both properties in nonlinear light-matter systems. In the following we demonstrate that resonance and geodesic evolution, though essential, are insufficient for robustness; stability is instead dynamically enabled by the geometric alignment of coherent rotation and dissipative contraction.

The remainder of this work is organized as follows. In Sec. II, the model is introduced and the main features of the dynamics are discussed, together with the Lindblad equation describing the interaction with a zero-temperature environment. In Sec. III, the entanglement dynamics is analyzed through the Negativity for different parameter regimes and a parametrized family of initial conditions, and geometric interpretations of the observed behavior are presented. Section IV constitutes the central part of this work, where robustness of the geometric phase is shown to be tied to the preservation of the geometric structure of the underlying unitary dynamics under dissipation. Finally, Sec. V summarizes and concludes the work.
%====================================================================

\section{Nonlinear Jaynes-Cummings Model}\label{sec:nonlinear_jcm}

We build upon the analysis of \cite{Viotti_2021} by including a Kerr-type nonlinear contribution in the cavity mode. Recent progress in superconducting circuits has emphasized the dual character of such nonlinearities: on the one hand, they can be deliberately engineered to provide qubit protection and enable parametric amplification \cite{Bhandari2025}; on the other, they may generate unwanted effects such as drive-induced transitions and additional dissipation channels that ultimately constrain device performance \cite{Yanik2026}. Motivated by this flexibility, we examine the Jaynes–Cummings model supplemented by a Kerr nonlinearity, describing a two-level atom coupled to a nonlinear single-mode cavity field. The Hamiltonian takes the form

\begin{equation}
\hat{H}=\frac{\Delta}{2}\hat{\sigma}_z+\chi \hat{n}^2
+g(\hat{\sigma}_+\hat{a}+\hat{\sigma}_-\hat{a}^\dagger).
\end{equation}

Here $\Delta=\omega_a-\omega_c$ denotes the atom–cavity detuning, $\chi$ is the strength of the Kerr nonlinearity, $g$ is the atom–field coupling constant, $\hat{\sigma}_\pm=\hat{\sigma}_x\mp i\hat{\sigma}_y$ are the atomic ladder operators, and $\hat{a}$ ($\hat{a}^\dagger$) represent the annihilation (creation) operators of the cavity mode.

This Hamiltonian follows from the application of the rotating-wave approximation to the Rabi Hamiltonian \cite{jc_original}, together with a transformation to the rotating frame $R=\exp{-iw_ct(\hat{a}^\dagger \hat{a} +\hat{\sigma}_z)}$. The validity of the RWA is ensured in the regime of weak atom–resonator coupling and near resonance. The parameters employed in the numerical simulations throughout this work lie well within this regime.

Within this frame, the detuning $\Delta$ determines the free evolution of the uncoupled atom–cavity system.

As in the linear Jaynes–Cummings model, the total number of excitations is conserved. The Hamiltonian therefore decomposes into invariant two-dimensional subspaces labeled by the excitation number $n$, each characterized by the block structure

\begin{equation}
\hat{H}^{(n)}=
\begin{pmatrix}
\frac{\Delta}{2}+\chi(n-1)^2 & g\sqrt{n} \\
g\sqrt{n} & -\frac{\Delta}{2}+\chi n^2
\end{pmatrix}.
\end{equation}

The Kerr contribution affects only the diagonal entries of each block. By expressing the diagonal part as $E_0(n) \mathbb{I}+\frac{\tilde\Delta(n)}{2} \hat\sigma_z$, we identify an excitation-dependent effective detuning $\tilde\Delta(n)=\Delta-\chi(2n-1)$.

In this way, the nonlinearity introduces a sector-dependent frequency shift. For closed-system dynamics, where the excitation number remains fixed, this amounts to a renormalization of the detuning within each invariant subspace.

Diagonalizing $\hat{H}^{(n)}$ leads to the dressed eigenstates

\begin{equation}\label{ec:autoestado kerr}
\begin{split}
    \ket{\psi_\pm^n}=\frac{1}{N_\pm}
\left[
\left(-\frac{\Delta}{2}+\chi\left(n-\frac{1}{2}\right)
\mp\frac{\Omega_{n,\chi}}{2}\right)\ket{g,n}
\right. \\ 
 + g\sqrt{n}\ket{e,n-1}
\bigg],
\end{split}
\end{equation}
where the generalized Rabi frequency is $\Omega_{n,\chi}=\sqrt{(\chi(2n-1)-\Delta)^2+4g^2n}$ and $N_\pm$ are normalization factors. The corresponding eigenenergies are

\begin{equation}\label{ec:autoenergia kerr}
E_\pm^n=\chi\left(n-\frac{1}{2}\right)^2
+\frac{\chi}{4}
\pm\frac{\Omega_{n,\chi}}{2}.
\end{equation}

The resonance condition is thus modified to $\Delta=\chi(2n-1)$, for which the effective detuning $\tilde{\Delta}(n)$ vanishes. Under this condition, the Hamiltonian in the $n$-th excitation sector becomes proportional to an effective $\tilde{\sigma}_x$ operator in the ${\ket{e,n-1},\ket{g,n}}$ basis, and the population dynamics is governed purely by coherent Rabi oscillations. For example, at resonance, the time evolution of the state $\ket{g,n}$ reduces to

\begin{equation}
\begin{split}
    \ket{g,n(t)}=
e^{-i\left[\chi\left(n-\frac{1}{2}\right)^2+\frac{\chi}{4}\right]t}
\bigg[
\cos\left(\frac{\Omega_{n,\chi}t}{2}\right)\ket{g,n}
\\-\sin\left(\frac{\Omega_{n,\chi}t}{2}\right)\ket{e,n-1}
\bigg],
\end{split}
\end{equation}
which explicitly displays Rabi oscillations with frequency $\Omega_{n,\chi}$. Nevertheless, several features of the model require closer examination, particularly the influence of the initial state and the associated entanglement dynamics.

To account for the coupling to a zero-temperature environment, we describe the open-system evolution through the Lindblad master equation for the system density matrix,
\begin{equation}\label{eq:lindblad_jcm}
     \hat{\dot\rho}(t) = -i[\hat{H},\hat{\rho}] +\gamma\mathcal{D}(\hat{a}) + p\mathcal{D}(\hat{\sigma}_-) + p_z\mathcal{D}(\hat{\sigma}_z),
\end{equation}
where the dissipator associated with a collapse operator $\hat O$ is defined as $\mathcal{D}(\hat O)=\frac{1}{2} \big( 2\hat O \rho(t) \hat O^{\dagger} - \{ \hat O^{\dagger}\hat O, \rho(t) \} \big)$. The commutator term generates coherent evolution under $\hat H$, while the remaining contributions describe the interaction with the environment. The term $\gamma\mathcal{D}(\hat{a})$ accounts for photon leakage due to cavity losses, with decay rate $\gamma$. The contribution $p\mathcal{D}(\hat{\sigma}_-)$ describes atomic energy relaxation at rate $p$ (at finite temperature, this would be accompanied by incoherent excitation processes), and $p_z\mathcal{D}(\hat{\sigma}_z)$ represents pure atomic dephasing with rate $p_z$.

At zero temperature, the environment induces only irreversible decay processes, progressively removing excitations from the system. This competes directly with the coherent atom–cavity exchange discussed above and has a strong impact on both entanglement generation and phase accumulation.

Restricting the Hilbert space to low-excitation sectors, the master equation reduces to a closed system of coupled differential equations for the relevant density-matrix elements. Explicitly, using the ordered basis $\mathbb{B}=\{\ket{g0},\ket{e0},\ket{g1},\ket{e1},\dots\}$
\begin{equation}
\begin{aligned}
\dot{\rho}_{00} &= p \rho_{11} + \gamma \rho_{22}, \\
\dot{\rho}_{11} &= -i g (\rho_{21} - \rho_{12}) - p \rho_{11}+\gamma \rho_{33}, \\
\dot{\rho}_{22} &= -i g (\rho_{12} - \rho_{21}) - \gamma \rho_{22}+2\gamma\rho_{44}+p\rho_{33}, \\
\dot{\rho}_{12} &= -i g (\rho_{22} - \rho_{11}) - i (\Delta-\chi) \rho_{12} - \frac{\gamma}{2} \rho_{12}-\frac{p}{2}\rho_{12} \\ &+\gamma\sqrt{2}\rho_{34}-2p_z\rho_{12}, \\
\dot \rho_{33} &= i\sqrt{2}g(\rho_{34}-\rho_{43})-(p+\gamma)\rho_{33} ,\\
\dot \rho_{34} &= -i\sqrt{2}g(\rho_{33}-\rho_{44})-i(\Delta-3\chi)\rho_{34}\\ &-(\frac{p}{2}+\frac{3\gamma}{2}+2p_z)\rho_{34} ,\\
\dot \rho_{44} &= -i\sqrt{2}g(\rho_{34}-\rho_{43})-2\gamma\rho_{44},
\end{aligned}
\end{equation}

This set of equations makes transparent how the Kerr nonlinearity alters not only the coherent oscillation frequencies, through the shifted detunings $\Delta-\chi$ for the $n=1$ and $\Delta-3\chi$ for the $n=2$ subspaces, but also indirectly reshapes the dissipative dynamics by modifying the resonance conditions between excitation sectors.

Given that our main focus is on the quantum-information aspects of the model, we confine our study to the low-excitation regime. In this regime, it is possible to characterize in detail both the dynamical creation and degradation of entanglement, as well as the geometric phase accumulated during the evolution.

As this work extends the results of \cite{Viotti_2021,Viotti_2024}, we refer the reader to those references for a comprehensive discussion of the dynamics and geometric phase under specific initial conditions. Here, our analysis proceeds in two stages. First, we examine the entanglement dynamics for different classes of initial states. We then identify a new feature that arises in the behavior of the geometric phase when considering geodesic evolution and dissipative dynamics as core elements of the robustness of the geometric phase.

In particular, we will show that the combined effect of resonance and initial condition, and environmental damping produces qualitative changes in both entanglement generation and in the robustness of the geometric phase.

\section{Entanglement in nonlinear JCM}\label{sec:entang_jcm}
We now analyze the entanglement dynamics of the model for different initial conditions. For simplicity, we restrict the discussion to the single-excitation subspace ($n=1$).
To quantify entanglement, we use Negativity \cite{Vidal_2002}, defined as
\begin{equation}
    \mathcal{N}(\rho)=\sum_{\lambda<0}|\lambda|=\frac{||\rho^{T_A}||_1-1}{2}
\end{equation}
where $||\cdot||_1$ denotes the trace norm and $\rho^{T_A}$ is the partial transpose of $\rho$ with respect to subsystem $A$. The sum runs over the negative eigenvalues $\lambda$ of $\rho^{T_A}$. Negativity is an entanglement monotone under local operations and classical communications (LOCC) \cite{Horodecki_2009}, and is straightforward to compute for mixed states.
Another option is to use Concurrence, for which the results are the same. For a detailed comparison between these measures we refer the reader to \cite{Verstraete_2001}.
Since the Hamiltonian decomposes into $2\times 2$ invariant blocks, the effective evolution within each excitation subspace admits a Bloch-sphere representation. In fact, the Hamiltonian restricted to the $n$-excitation sector generates an SU(2) rotation with axis

\begin{equation}\label{eq:hamiltonian_direction}
\vec{n}=g\sqrt{n}\vec{e_x}+\frac{\Delta-\chi(2n-1)}{2}\vec{e_z},
\end{equation}
where $\vec{e}_j$ denotes the unit vector along direction $j$ of the Bloch sphere, with its basis vectors $\mathcal{B}=\{ \ket{g1},\ket{e0} \}$.

The dynamics in this representation is particularly transparent. It is easy to see that the atom–cavity entanglement is maximal for states lying in the equatorial plane ($z=0$) of the Bloch sphere, namely\begin{equation}
    \ket{\psi_{MES}}=\frac{1}{\sqrt{2}}(\ket{e0}+e^{i\phi}\ket{g1}),
\end{equation} 
with $\phi\in[0,2\pi]$ the azimuthal angle.
To study the dependence on the initial condition, we first impose the resonance condition in the single-excitation sector, $n=1$, i.e., $\Delta=\chi$, so that the Hamiltonian generates rotations around the $x$-axis of the Bloch sphere. We consider the general pure initial state
\begin{equation}\label{eq:initial_condition_jcm}
    \ket{\psi(0)}=\cos\frac{\theta_0}{2}\ket{e0}+e^{i\phi_0}\sin\frac{\theta_0}{2}\ket{g1}
\end{equation}
and fix $\phi_0=0$, varying $\theta_0\in[0,\pi/2]$.

\begin{figure*}[ht!]
    \centering
    \includegraphics[width=\textwidth]{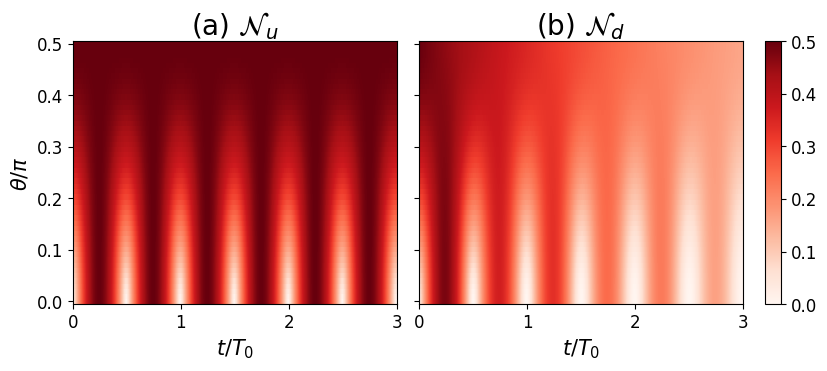}
    \caption{Negativity dynamics in time in resonance condition $\Delta=\chi$ for different initial conditions for (a) closed systems and (b) open system evolution.}
    \label{fig:negatividad_tita_jcm}
\end{figure*}

As expected, the trajectories correspond to rotations about the $x$-axis. For $\theta_0=0$, the initial state lies in the $yz$-plane and the trajectory follows a great circle, i.e., a geodesic of the Bloch sphere. As $\theta_0$ approaches $\pi/2$, the initial state becomes increasingly aligned with the rotation axis, and the corresponding trajectories shrink until, for $\theta_0=\pi/2$, the state becomes an eigenvector of the Hamiltonian and remains stationary.
Figure \ref{fig:negatividad_tita_jcm} displays the time evolution of the Negativity for (a) a closed system and (b) an open system with cavity loss rate $\gamma=0.1g$ and pure dephasing rate $p_z=0.01g$. In both cases, the largest oscillation amplitude occurs for $\theta_0=0$. As $\theta_0$ increases (i.e., as the initial state approaches the rotation axis), the oscillation amplitude decreases while the mean entanglement increases.
This behavior admits a simple geometric interpretation: since the dynamics is a rotation around the $x$-axis, states in the $yz$-plane periodically cross the equator—where entanglement is maximal—and the poles—where the state is separable. When the initial state is closer to the rotation axis, its trajectory remains closer to the equatorial region, thereby increasing the average entanglement while reducing oscillation amplitude.
\begin{figure*}[ht!]
    \centering
    \includegraphics[width=\textwidth]{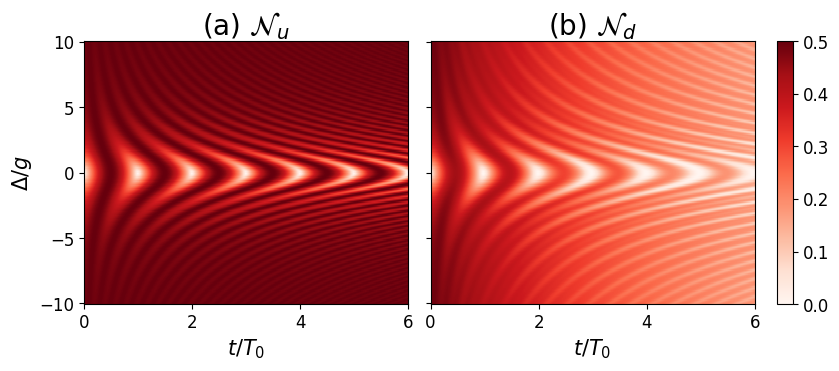}
    \caption{Negativity dynamics in time for different values of $\Delta$ and $\chi=0$. The initial condition is perpendicular to de direction of the Hamiltonian; for (a) closed systems and (b) open system evolution. Changing $\chi$ only shifts this figure so that the symetry axis corresponds to resonance $\Delta=\chi$.}
    \label{fig:negatividad_perpendicular_jcm}
\end{figure*}
In the open-system case, the environment progressively degrades entanglement. While in the closed system the oscillations preserve their amplitude, in the dissipative case the oscillations are damped and the Negativity eventually decays toward zero.

We now relax the resonance condition and allow arbitrary values of $\Delta$ and $\chi$. To explore the influence of the rotation axis, we select a particular class of initial states. In previous works \cite{Viotti_2021}, robustness of certain quantum properties was associated with trajectories that follow geodesics in state space. Motivated by this observation, we choose for each pair $(\Delta,\chi)$ an initial state perpendicular to the Hamiltonian direction:

\begin{equation}\label{eq:perpendicular_state}
\begin{split}
    \ket{\psi_\perp}&=\cos\frac{\theta_n+\pi/2}{2}\ket{e,n-1}+\sin\frac{\theta_n+\pi/2}{2}\ket{g,n}\\
    &=\frac{1}{\sqrt{2}}\left(\ket{+,\hat n}-\ket{-,\hat n}\right),
\end{split}
\end{equation}
where $\theta_n$ defines the polar angle of the Hamiltonian direction $\hat n$ in Eq. \eqref{eq:hamiltonian_direction}. By construction, this state is perpendicular to the rotation axis, so that the unitary evolution traces a great circle on the Bloch sphere, i.e., a geodesic.

Figure \ref{fig:negatividad_perpendicular_jcm} shows the corresponding entanglement dynamics. In the particular case when $\Delta=\chi$, the resonance condition is fulfilled (for $n=1$) and Negativity shows maximal oscillation amplitude. Then, the rotation axis is aligned with the $x$-direction, and the perpendicular initial state lies in the $yz$-plane. The trajectory therefore connects the north and south poles (separable states) through the equator (maximally entangled states), yielding full entanglement oscillations.

As $|\Delta|$ increases, the Hamiltonian axis tilts toward the $z$-direction. A state perpendicular to this axis approaches the equatorial ($xy$) plane, where entanglement is maximal. Consequently, the dynamics becomes confined to a region of high entanglement.

Changing $\chi$ only shifts this figure so that the symetry axis corresponds to resonance $\Delta=\chi$. Nonlinearities do not change the dynamics in this case, since we only have $n=1$ and therefore the excitation number dependence on the frequencies does not play a role here.

\section{Geometric phase in JCM}\label{sec:gp_jcm}
In this section we employ two central results from geometric phase (GP) theory \cite{Berry1984,Bhandari1988,Mukunda1993,Pancha1956}, namely the kinematic formulations for general unitary and non-unitary evolutions.

For unitary dynamics of a pure state $\ket{\psi(t)}$, the geometric phase associated with a path $C$ in projective Hilbert space is given by \cite{Mukunda1993}
\begin{equation}\label{eq:phi_u}
    \phi_u(t)=\arg \braket{\psi(t_0)}{\psi(t)} - \Im \int_{t_0}^tdt'\braket{\psi(t')}{\dot \psi(t')}.
\end{equation}
This expression separates the total phase into a Pancharatnam \cite{Pancha1956} overlap term and a dynamical contribution, isolating the purely geometric component associated with the trajectory in state space.

For non-unitary evolution of a mixed state $\rho(t)=\sum_k \omega_k(t)\ket{\psi_k(t)}\bra{\psi_k(t)}$, the kinematic approach yields \cite{Tong2004}

\begin{eqnarray}
    \phi_g(t)&=\arg \Big( \sum_{k=1}^{N} \sqrt{\omega_k(t_0)\omega_k(t)} \braket{\psi_k(t_0)}{\psi_k(t)}  \nonumber \\
     &\times  e^{-\int_{t_0}^{T}dt'\braket{\psi_k(t')}{\dot\psi_k(t')}} \Big)\label{eq:phi_g} .
\end{eqnarray}

This formulation generalizes Berry’s original construction to arbitrary open-system evolutions, provided that the eigenvalues remain non-degenerate and the spectral decomposition evolves smoothly in time. It has been widely used to study different systems \cite{LV2010,TLSLV2010,LV2013,LV2015}, and theory has been developed to understand how it is affected by decoherence \cite{LV2006}. In the case of an initially pure state, only a single eigenvalue is equal to one at $t=t_0$, so that the geometric phase can be associated with the evolution of the corresponding eigenvector of the density matrix. In that case, the expresion \eqref{eq:phi_g} reduces to
\begin{equation}\label{eq:phi_plus}
    \phi_g(t)=\arg \braket{\psi(t_0)}{\psi_+(t)} - \Im \int_{t_0}^tdt'\braket{\psi_+(t')}{\dot \psi_+(t')},
\end{equation}
where $\ket{\psi_+(t)}$ represents the eigenstate of the density matrix whose eiganvalue was one at initial time. In this context, the geometric phase could be interpreted as the one accumulated by this pure vector, even though the physical state is a statistical ensemble of vectors.

For an initial study of the geometric phase in this model we refer to \cite{Viotti_2021}, where the Jaynes-Cummings model has been already studied and later applied to a transmon in a superconducting circuit \cite{Viotti_2024}. There, it was shown that under the resonance condition $\Delta=\chi=0$, the geometric phase exhibits robustness against environmental effects.

The physical origin of this robustness can be understood as follows. For an initially pure state, the geometric phase of the open system can be identified with the phase accumulated by the eigenvector of the density matrix whose initial eigenvalue is unity. As the system evolves, both this eigenvalue and eigenvector change continuously, and the latter traces a trajectory in the Bloch sphere. It is shown that in the resonant case, the trajectory of the corresponding eigenvector coincides with the trajectory in the case of closed system dynamics: a geodesic of the Bloch sphere. Therefore, the geometric phase is equal in both the closed and open system case and has a characteristic step like dynamics, each time it crosses over the opposite pole the GP makes a $\pi$-jump.  

In the zero-temperature case considered here, the environment eventually drives the system toward its ground state. Consequently, the eigenvalue associated with the initially pure component decays to zero, and the notion of geometric phase loses physical meaning in the long-time limit, as the state becomes effectively mixed. Nevertheless, for intermediate times—before complete relaxation—the geometric phase remains well defined and experimentally accessible \cite{pechal}.

\subsection{Resonance and dependence on initial condition}

We now extend the analysis to the nonlinear case under the resonance condition for $n=1$, namely $\Delta=\chi\neq0$, and investigate the role of the initial state. Following the procedure of Sec. \ref{sec:entang_jcm}, we vary $\theta_0\in[0,2\pi]$ in Eq. \eqref{eq:initial_condition_jcm}, fixing $\phi_0=0$.

\begin{figure}
    \centering
    \includegraphics[width=0.5\textwidth]{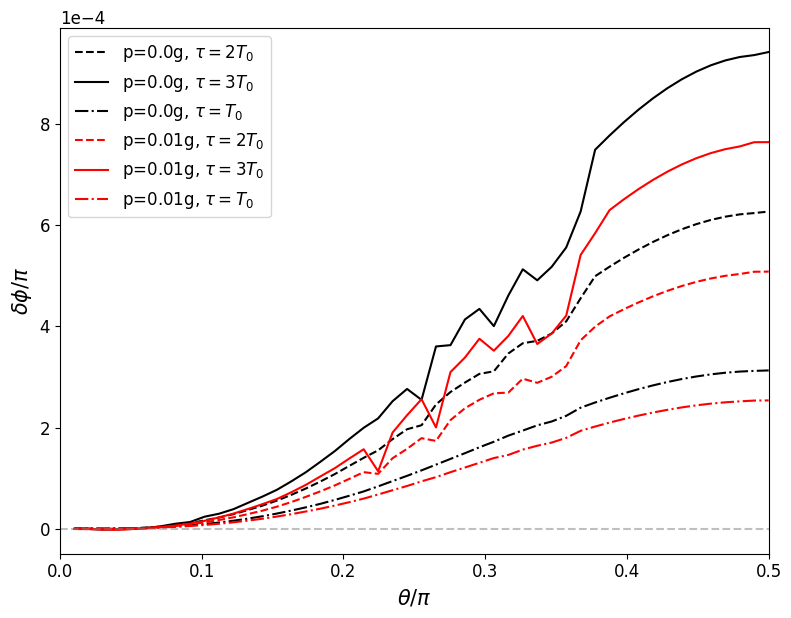}
    \caption{Difference of GP between closed and open system after a time $\tau$ when varying initial condition direction $\theta_0$ while maintaining resonance $\Delta=\chi$ so that the Hamiltonian represents rotation around the $x$-axis of the Bloch sphere.}
    \label{fig:robustness_initial_condition}
\end{figure}

To quantify robustness, we compute the difference $\delta\phi=\phi_g-\phi_u$ after a fixed evolution time $\tau=m\cdot \frac{2\pi}{\Omega}$, where $\phi_g$ corresponds to the open-system geometric phase and $\phi_u$ to the unitary case, and $m\in\mathbb{N}$. The quantity $\delta\phi$ thus provides a measure of the sensitivity of the geometric phase to environmental decoherence. Figure \ref{fig:robustness_initial_condition} shows the dependence of $\delta\phi$ on the initial condition. The results indicate that satisfying the resonance condition alone is not sufficient: robustness is achieved only when the trajectory follows a geodesic in the Bloch sphere. Initial states outside the maximal circle exhibit significant deviations between $\phi_g$ and $\phi_u$. 

This confirms that geometric robustness is not merely a spectral property of the Hamiltonian, but a global property of the path traced in state space.

\subsection{Beyond resonance: perpendicular initial conditions}

Motivated by the previous observations, we now abandon the resonance condition and allow arbitrary values of $\Delta$ and $\chi$. The Hamiltonian then generates rotations around the general axis defined in Eq. \eqref{eq:hamiltonian_direction}.

\begin{figure}
    \centering
    \includegraphics[width=0.5\textwidth]{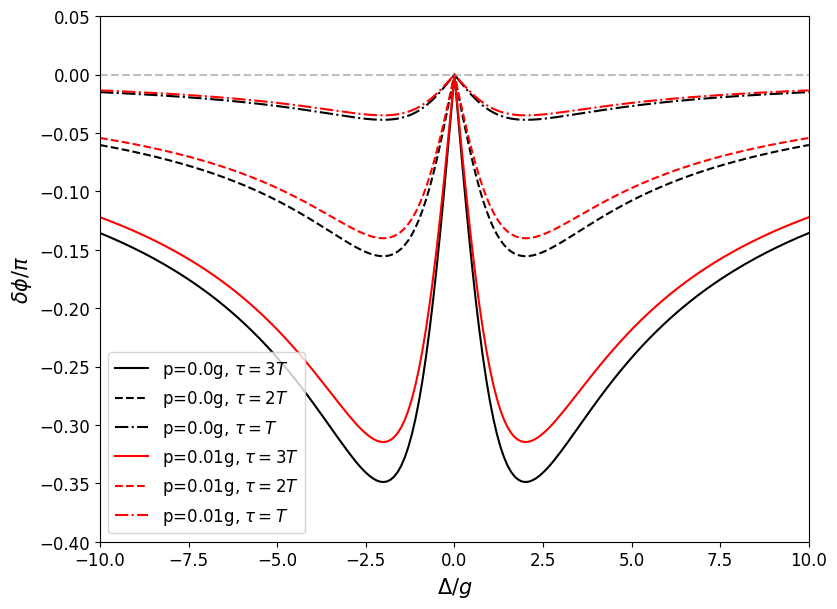}
    \caption{Difference of GP between closed and open system after a time $\tau$ while varying $\Delta$ (off-resonance) but picking $\theta_0$ so that the initial condition is perpendicular to the Hamiltonian direction.}
    \label{fig:robustness_perpendicular}
\end{figure}

Choosing the initial state perpendicular to this axis, as in Eq. \eqref{eq:perpendicular_state}, the closed-system evolution again traces a geodesic. The natural question is whether the GP in such trajectories is always robust. Figure \ref{fig:robustness_perpendicular} shows that this is not the case. Even when the closed-system path is geodesic, robustness of the geometric phase is not guaranteed. Therefore, that the unitary trajectory follows a geodesic path is a necessary but not sufficient condition for robustness. For the robustness to be true, not only does the system need to evolve along a geodesic, but also the decoherence process should have some specific dynamical properties, achieved in this case under resonance condition. 

It is worth noting that in this case, the measurment times $\tau=T,2T$ and $3T$ are proportional to the oscilation period $T$ of the system. As $\Delta$ increases, the period decreases. Consequently, the effect is stronger than it may appear at first glance: although the measurements are performed at shorter times for larger $\Delta$, the difference $\delta\phi$ nonetheless increases, showing real protection to decoherence when in resonance.

\begin{figure}
    \centering
    \begin{subfigure}{0.49\linewidth}
        \centering
        \includegraphics[width=\textwidth]{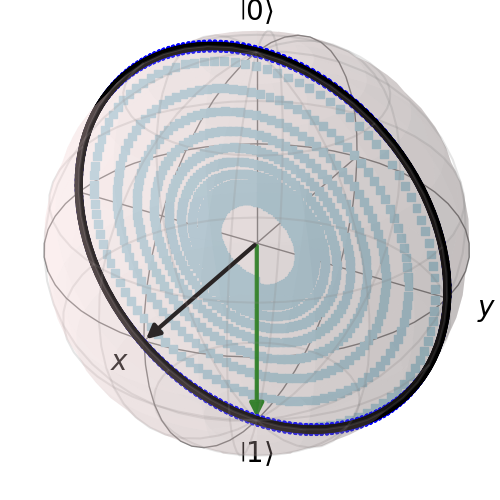}
        \caption{On resonance}
        \label{fig:bloch_resonance}
    \end{subfigure}
    \hfill
    \begin{subfigure}{0.49\linewidth}
        \centering
        \includegraphics[width=\textwidth]{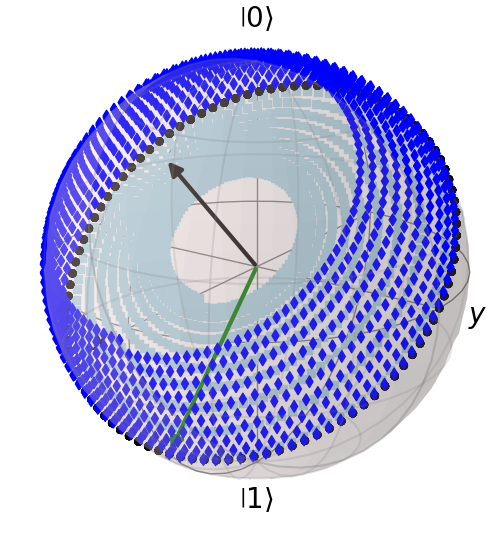}
        \caption{Off resonance}
        \label{fig:bloch_off_resonance}
    \end{subfigure}
       \caption{Trajectories on Bloch sphere for perpendicular initial conditions. The black arrow represents the Hamiltonian (i.e. the rotation axis) and the green arrow the initial condition. In black is the unitary trajectory, and in light-blue the projection of the density matrix in this space for a Lindbladian dynamic. Finally, in blue, is the eigenvector $\ket{\psi_+(t)}$ which accumulates GP in the later case. Here, $\ket{0}\equiv\ket{e0}$ and $\ket{1}\equiv\ket{g1}$.}
       \label{fig:bloch_trayectories}
\end{figure}

To clarify this behavior, we analyze the Bloch-sphere representation of the open-system dynamics projected onto the $n=1$ subspace. Dissipation causes the Bloch vector to spiral toward the center of the sphere. However, the geometry of this spiral depends crucially on the value of $\tilde\Delta(n)$. 
As shown in Fig. \ref{fig:bloch_resonance}, under resonance the dissipative spiral lies in the same plane as the unitary geodesic. Consequently, the dominant eigenvector of the density matrix follows precisely the same trajectory as in the closed system, and the geometric phase remains unchanged.

In contrast, Fig. \ref{fig:bloch_off_resonance} shows that, off resonance, the dissipative contraction is not contained in a tilted plane. Although the closed-system evolution is still geodesic, the eigenvector associated with the largest eigenvalue no longer follows the same path as in the unitary case. As a result, the geometric phase accumulated in the open system deviates from the unitary value.

This result is significant for two reasons. First, it deepens our understanding of the interplay between Kerr-induced detuning and decoherence in the Jaynes–Cummings model. Second, it shows that robustness of the geometric phase is not universal: it depends on an interplay between the Hamiltonian parameters and the structure of the dissipative dynamics.

The analysis above reveals a subtle but important point: the robustness of the geometric phase in the nonlinear Jaynes–Cummings model is not determined solely by the geometric character of the unitary trajectory. While geodesic evolution is a necessary ingredient, it is not sufficient. What ultimately governs robustness is the interplay between coherent rotation and dissipative contraction in state space. When the Lindblad dynamics preserves the plane of the unitary trajectory—as occurs under resonance—the dominant eigenvector of the density matrix follows the same path as in the closed system, and the geometric phase remains unchanged. Off resonance, however, the dissipative process takes the contraction off of the plane, modifying the eigenvector trajectory and thus altering the accumulated geometric phase. This observation provides a refined perspective on decoherence in cavity QED systems: environmental effects do not merely suppress quantum features, but reshape the geometry of quantum trajectories in state space. The geometric phase is robust only when this reshaping preserves the structure of the underlying coherent path. From a broader viewpoint, this suggests that geometric protection mechanisms in open quantum systems should be understood in terms of dynamical symmetries of the full Liouvillian evolution. Identifying such symmetries may provide a systematic route to engineering robustness beyond specific resonance conditions. 
Importantly, these geometric considerations are not restricted to the single-atom Jaynes–Cummings model. The single-atom analysis presented here serves as a minimal geometric framework that anticipates the more intricate collective effects appearing in more complex systems, as for example in a multi-emitter generalization of this problem.
\section{Conclusions}\label{sec:conclusions}

The geometric phase and entanglement dynamics of the nonlinear Jaynes–Cummings model with Kerr nonlinearities and dissipation are analyzed, focusing on the mechanisms that determine the robustness of geometric phases in hybrid atom–cavity systems.

It was shown that Kerr nonlinearities mainly renormalize the resonance condition through photon-number-dependent frequency shifts. In this regime, resonance conditions lead to highly regular and coherent oscillations of both population and entanglement. However, the analysis of open-system dynamics reveals that the existence of quasi-geodesic unitary trajectories is not, by itself, sufficient to guarantee robustness of the geometric phase under decoherence. Instead, robustness emerges only when the interplay between initial conditions, dissipation channels, and resonance structure dynamically preserves some properties of the unitary trajectory, in particular the density matrix eigenstate traces the same path as the unitary state. These results highlight the fundamental role of atom--cavity hybridization in stabilizing geometric properties of the evolution.

A key conclusion of this study is that the robustness of the geometric phase in open hybrid quantum systems is dynamically enabled and not purely geometric in origin. Two ingredients appear to be particularly relevant: (i) resonance between bare configurations that promotes strong and coherent dressed-state hybridization, and (ii) the combined action of initial state preparation and dissipation channels, which determines whether the system remains in the coherent sector that supports stable geometric phase accumulation.

These results open the possibility of engineering dynamical protocols where geometric robustness is actively maintained through tailored system–environment interactions.

\section*{Acknowledgments}
We thank Consejo Nacional de Investigaciones Científicas y Técnicas (CONICET) and Universidad de Buenos Aires (UBA) for their financial support.

\medskip
\bibliography{zbibliografia.bib}

\end{document}